\documentclass[twoside]{article}
\usepackage{fleqn,espcrc2}
\usepackage{amsfonts}

\newcommand{\tr}{{\rm tr}}
\newcommand{\ad}{{\rm ad}}

\newcommand{\om}{\omega}
\newcommand{\Om}{\Omega}
\newcommand{\de}{\delta}

\newcommand{\te}{\theta}
\newcommand{\vth}{\vartheta}

\newcommand{\La}{\Lambda}
\newcommand{\D}{\Delta}
\newcommand{\ve}{\varepsilon}

\newcommand{\vf}{\varphi}
\newcommand{\G}{\Gamma}

\newcommand{\ze}{\zeta}

\def\bfe{{\bf e}}

\def\bv{{\bf v}}
\def\cS{{\cal S}}

\def\mC{{\mathbb C}}
\def\mZ{{\mathbb Z}}
\def\mR{{\mathbb R}}
\def\mN{{\mathbb N}}
\def\rot{{\rm curl}}
\newcommand{\ar}[2]{\!\left[\!\begin{array}{c} {#1}\\{#2}\end{array}\!\right]}

\newcommand{\beq}[1]{\begin{equation}\label{#1}}
\newcommand{\eq}{\end{equation}}
\newcommand{\beqn}[1]{\begin{eqnarray}\label{#1}}
\newcommand{\eqn}{\end{eqnarray}}
\newcommand{\p}{\partial}
\newcommand{\di}{{\rm diag}}
\newcommand{\oh}{\frac{1}{2}}

\newcommand{\GLN}{{\rm GL}(N,{\mathbb C})}

\newcommand{\gln}{{\rm gl}(N, {\mathbb C})}

\def\f1#1{\frac{1}{#1}}

\newcommand{\bp}{\bar{\partial}}
\newcommand{\bz}{\bar{z}}

\def\frak{\mathfrak}
\def\gg{{\frak g}}
\def\gJ{{\frak J}}

\newcommand{\AmS}{{\protect\the\textfont2
  A\kern-.1667em\lower.5ex\hbox{M}\kern-.125emS}}

\title{Integrable tops and non-commutative torus}

\author{M. Olshanetsky
\address{Institute of Theoretical and Experimental Physics, Moscow, Russia,\\
 Max Planck Institute of Mathematics,  Bonn, Germany.\\
{\em e-mail olshanet@gate.itep.ru}}}

\begin{document}

%\begin{flushright{
%{\it very  preliminary draft}
%\end{flushright{
\begin{abstract}
We consider the hydrodynamics of the ideal fluid on a
2-torus and its Moyal deformations. The both type of
equations have the form of the Euler-Arnold tops.
The Laplace operator  plays the role of the inertia-tensor.
It is known that 2-d hydrodynamics is non-integrable.
After replacing of the Laplace operator by  a distinguish 
pseudo-differential operator the deformed system becomes integrable.
It is an infinite rank Hitchin system over an elliptic curve with
transition functions from the group of the non-commutative torus.
In the classical limit we obtain an integrable analog of the hydrodynamics on
a torus with the inertia-tensor operator $\bar\partial^2$ instead
of the conventional Laplace operator $\partial\bar\partial$.
\vspace{1pc}
\end{abstract}

\maketitle

%\thanks{Present address: Max Planck Institute of Mathematics,     
%Vivatsgasse 7, D-53111 Bonn, Germany.\\
%The work is supported in part  by grants
%RFBR-00-02-16530, 00-15-96557 for support of scientific schools,
%and INTAS-99-01782.}


\section{Introduction}
\setcounter{equation}{0}

The Euler-Arnold tops (EAT) are Hamiltonian systems defined
on the coadjoint orbits of groups \cite{Ar}. Particular examples of
such systems are the Euler top related to SO(3), its SO($N$) generalization
\cite{Man} 
and  the hydrodynamic of the ideal incompressible
fluid on a space $M$. 
The corresponding group of the latter system is SDiff$(M)$.
We consider here the case $\dim(M)=2$ and then restrict $M$ 
to be a torus $T^2$.
 
EAT are completely determined by their Hamiltonians,
since the Poisson structure is fixed to be related to
 the Kirillov-Kostant form on the coadjoint orbits.
The Hamiltonians are determined by the inertia-tensor operator $\gJ$
mapping the Lie algebra $\gg$ to the Lie coalgebra $\gg^*$.
 Special choices of $\gJ$ lead to completely integrable systems 
(see review \cite{DKN}). In the case of
the 2d hydrodynamics $\gJ$ is the Laplace operator and it
turns out that the theory is non-integrable \cite{Z}.
The goal of this paper are integrable models related to \mbox{SDiff}.
Some integrable models related to SDiff were considered in
\cite{SV,GKR,TT,T}. 

These type of models can be described as the classical limit
of integrable models when the commutators in the Lax equations 
are replaced by the Poisson brackets. This approach was
proposed in Ref.\cite{LM}, where the dispersionless KP hierarchy
was constructed and later developed in numerous
publications (see review \cite{TT1}).

Here we use the same strategy defining an integrable system on
the non-commutative torus (NCT) and then taking the classical
limit to SDiff$(T^2)$.
The starting point in our construction is the integrable $\GLN$
EAT introduced in
Ref.\cite{STSR,LOZ}. Their  inertia-tensor operators depend on the module
$\tau,~Im\tau>0$ of an elliptic curve. This curve is the basic
spectral curve in the Hitchin description of the model. 
We consider here a special limit $N\to\infty$ of $\GLN$ that leads to
 the group $G_\te$ of NCT, where $\te$ is the Planck 
constant.\footnote{For Manakov's top $\lim N\to\infty$ 
was considered in Ref.\cite{War}.}
 The group 
$G_\te$ is defined as the
set of invertible elements of the  NCT algebra ${\cal A}_\te$. 
It can be embedded in GL$(\infty)$ and in this way $G_\te$
can be described as a special limit of $\GLN$. We define  
EAT related to $G_\te$ depending on a parameter $\tau$.
Then, we construct the Lax operator with the spectral parameter
on an elliptic curve with the same parameter $\tau$.

In the classical limit $\te\to 0$, $G_\te\to$SDiff$(T^2)$ and the 
inertia-tensor operator $\gJ$ takes the form $\bp^2$. The conservation
laws survive in this limit while commutators in the Lax hierarchy
become the Poisson brackets. It turns out that
 the classical limit is essentially the same
as the rational limit of the basic elliptic curve, such that the product
of the Planck constant $\te$ and the half periods of the basic curve
are constant. In this way in the classical limit NCT
 becomes dual to the basic elliptic curve.

\section{The Lie algebra of the non-commutative torus}
\setcounter{equation}{0}

Here we reproduce some basic results about NCT and
related to it Lie algebra $Sin_\te$ \cite{FFZ}.

{\sl 1. Non-commutative torus.}

Quantum torus ${\cal A}_\te$ is the unital algebra with
 two generators $(U_1,U_2)$ that satisfy the relation
\beq{3.1}
U_1U_2=\bfe(\te) U_2U_1,~\bfe(\te)=e^{ 2\pi i \te},~ \te\in[0,1)\,.
\eq

Elements of ${\cal A}_\te$ are the double sums
$$
{\cal A}_\te=
\{x=\sum_{m,n\in{\mathbb Z}}a_{mn}U_1^mU_2^n,~a_{mn}\in\mathbb C\}\,,
$$
where $a_{mn}$ are rapidly decreasing on the lattice ${\mathbb Z}^2$
\beq{3.1a}
{\rm sup}_{m,n\in\mZ}(1+m^2+n^2)^k|a_{mn}|^2<\infty,\,
{\rm for}~k\in \mN\,.
\eq
The trace functional $\tr(x)$ on ${\cal A}_\te$ is defined as
$\tr(x)=a_{00}$. It is positive definite
 and satisfies the evident identities
$$
\tr(x^*x)=\sum_{m,n\in{\mathbb Z}}|a_{mn}|^2\geq 0\,,
$$
$$
\tr(1)=1,~~\tr(xy)=\tr(yx\,,
$$
where the star involution  acts as \mbox{ 
$U_a^*=U_a^{-1},~a_{mn}^*=\bar{a}_{mn}$.}
The relation with the commutative algebra of smooth functions
on the two dimensional torus 
\beq{T}
T^2=\{\mR^2/\mZ\oplus\mZ\}\,\sim \,\{0<x\leq 1,\,0<y\leq 1\}.
\eq
comes from the identification
\beq{3.2}
U_1\to\bfe(x),~~U_2\to\bfe(y),
\eq
and the multiplication on $T^2$ becomes the Moyal multiplication:
\beq{3.3}
(f*g)(x,y):=fg+
\eq
$$
\sum_{n=1}^\infty\frac{(\pi\te)^n}{n!}
\ve_{r_1s_1}\ldots\ve_{r_ns_n}(\p^n_{r_1\ldots r_n}f)
(\p^n_{s_1\ldots s_n}g).
$$
Another representation with the operators,
that acts on the space of functions on $\mR/\mZ$, has the form
\beq{3.4}
U_1\to\exp(i\vf),~U_2\to\bfe(\te\p_\vf)\,.
\eq
Finally, we can identify $U_1,U_2$ with matrices in GL$(\infty)$
\beq{3.5}
U_1\to Q,~U_2\to \La\,,
\eq
where $Q$ and $\La$ are defined as
$$
Q=\di (\bfe(j\te)),~\La=E_{j,j+1},~j\in\mZ\,.
$$
The trace functional in the Moyal identification (\ref{3.2})
is the integral
\beq{3.6}
\tr f=\int_{T^2}fdxdy=f_{00}\,.
\eq

{\sl 2. Sin-algebra}

Define the following quadratic combinations of the generators
\beq{3.10}
T_{mn}=\frac{i}{2\pi\te}\bfe
\left(\frac{mn}{2}\te
\right)U_1^mU_2^n\,.
\eq
Their commutator has the form of the sin-algebra 
\beq{3.11}
[T_{mn},T_{m'n'}]=
\eq
$$
=\f1{\pi\te}\sin\pi\te(mn'-m'n)T_{m+m',n+n'}\,.
$$
We denote  $Sin_\te$ the Lie algebra of the series
\beq{3.12}
\psi=\sum_{mn}\psi_{mn}T_{mn}
\eq
with rapidly decreasing coefficients (\ref{3.1a}) and the commutator
(\ref{3.11}).

In the Moyal representation (\ref{3.3}) the commutator has the form
$$
[f(x,y),g(x,y)]=\{f,g\}^*\,,
$$
where
\beq{3.14}
\{f,g\}^*=\f1{\te}(f*g-g*f)\,.
\eq

The algebra $Sin_\te$ has a central extension $\widehat{Sin}_\te$
that comes from the one-cocycles of  ${\cal A}_\te$. The corresponding 
additional term in (\ref{3.11}) has the form
\beq{3.13}
(am+bn)\de_{m,-m'}\de_{n,-n'}\,.
\eq

Let $\te$ be a rational number $\te=p/N$, where $p,N\in\mN$ with
gcd$(p,N)=1$.
Then $Sin_\te$ has an ideal 
\beq{I1}
I_N=\{T_{mn}-T_{m+Nl,n+Nk},\,k,l\in\mZ\},
\eq
and the factor $Sin_{p/N}/I_N$ for odd $N$ is isomorphic to $\GLN$.
Similarly, we can consider the ideal 
\beq{I2}
\hat{I}_N=\{T_{mn}-T_{m+Nl,n},\,l\in\mZ\},
\eq
in $\widehat{Sin}_\te$. The factor $\widehat{Sin}_\te/\hat{I}_N$ 
is isomorphic to 
the central extended loop algebra $\hat{L}(\gln)$.

\section{2d-hydrodynamics on ${\cal A}_\te$}
\setcounter{equation}{0}

{\sl 1. 2-d hydrodynamics}

Let $\bv=(V_x,V_y)$ be the velocity vector of the ideal incompressible
fluid on a compact manifold $M$ (dim$(M)=2$, div$\bv=0$).
\footnote{For simplicity we assume that the measure on $M$ is $dx\wedge dy$,
though all expressions can be written in a covariant way.}
The Euler equation for 2d hydrodynamics   takes
the form \cite{Ar}
\beq{Ar1}
\p_t\rot \bv=\rot[\bv,\rot \bv]\,, 
\eq
where $\rot \bv=\p_xV_y-\p_yV_x$
is the vorticity of the vector field $\bv$.

Let $\psi(x,y)$ be the stream function. It is
 the Hamiltonian generating the vector field $\bv$ 
$$
i_\bv dx\wedge dy=d\psi\,.
$$
In other words \mbox{
$V_x=-\p_y\psi,~~V_y=\p_x\psi$.} 
There is a isomorphisms of the Lie algebra of vector fields div$\bv=0$ on
 $M$ and the Poisson
algebra of the stream functions $\gg=\{\psi\}$ on $M$ defined up to constants 
$$
\{\psi_{\bv_1},\psi_{\bv_2}\}=\psi_{[\bv_1,\bv_2]}
$$

 Let $\gg^*$ be the dual space of 
distributions on $M$.
The vorticity $\cS=\rot \bv$ 
of the vector field $\bv$ 
 $$
\cS=-\Delta\psi
$$
can be considered as an element from $\gg^*$.
 The Euler equation (\ref{Ar1})  for $\cS$ takes the form
\beq{Ar2}
\p_t\cS=\{\cS,\psi\},~{\rm or}~~\p_t\cS=\{\cS,\Delta^{-1}\cS\}\,.
\eq

We can look on  (\ref{Ar2}) as the Euler-Arnold equation 
for the rigid top related
to the Lie algebra $\gg$, where the Laplace operator is the map
$$
\Delta \,: \gg\to\gg^*
$$
that plays the role of the inertia-tensor. The phase space of the system
is a coadjoint orbit of the group of the canonical transformations
 SDiff$(M)=\exp\{\psi,\cdot \}$. 
The equation (\ref{Ar2}) takes the form
\beq{Ar3}
\p_t{\cal S}=\ad^*_{\nabla H}{\cal S}\,,
\eq
where ${\nabla H}=\frac{\de H}{\de \cS}=\psi$ 
is the variation of the Hamiltonian
\beq{Ar4}
H=\oh\int_M\cS\Delta^{-1}\cS=\int_M\psi\Delta\psi\,.
\eq
There is infinite set of Casimirs defining the coadjoint
orbits
\beq{Ca}
C_k=\int_M\cS^k\,.
\eq
Certainly, these integrals cannot provide the integrability 
of the system \cite{Z}.

Consider a particular case, when $M$ is a two-dimensional torus (\ref{T}).
In terms of the Fourier modes $s_{mn}$ of the vorticity
$$
\cS=\sum_{mn}s_{mn}\bfe(mx+ny)
$$
the Hamiltonian (\ref{Ar4}) takes the form
\beq{Ar6}
H=-\f1{8\pi^2}\sum_{mn}\f1{m^2+n^2}s_{mn}s_{-m,-n}\,,
\eq
and we come to the equation 
\beq{Ar5}
\p_ts_{mn}=-\f1{8\pi^2}\sum_{j,k}\frac
{jn-km}{j^2+k^2}s_{jk}s_{m-j,n-k}\,.
\eq

{\sl 2. 2d hydrodynamics on non-commutative torus}.

We can consider the similar construction by replacing the
Poisson brackets by the Moyal brackets (\ref{3.14}) \cite{Ze,HOT}.
Introduce the vorticity $\cS$ as an element of $Sin^*_\te$
\beq{4.00}
\cS=\sum_{mn}s_{mn}T_{mn}.
\eq
The equation (\ref{Ar2}) takes the form
$$
\p_t\cS= \{\cS,\Delta^{-1}\cS\}^*\,,
$$
or for the Fourier modes 
\beq{Ar7}
\p_ts_{mn}=
\eq
$$
-\f1{8\pi^3\te}\sum_{j,k}\frac
{\sin\pi\te(jn-km)}{j^2+k^2}s_{jk}s_{m-j,n-k\,}\,.
$$
This system is the Euler-Arnold top on the group $G_\te$ of
invertible elements of ${\cal A}_\te$ and the coadjoint orbits
are defined by the same Casimirs (\ref{Ca}) as for SDiff$(T^2)$.

\section{ Elliptic rotator on ${\cal A}_\te$}
\setcounter{equation}{0}

Let $\wp$ be the Weierstrass function depending on the modular parameter
$\tau$, Im$\tau>0$.
\beq{4.0}
\wp(u;\tau)=\f1{u^2}+
\eq
$$
+\sum'_{j,k}\left(
\f1{(j+k\tau+u)^2}-\f1{(j+k\tau)^2}\right)\,.
$$
We replace the inverse inertia-tensor $\Delta^{-1}$ of the hydrodynamics
on the pseudodifferential operator $\gJ^{-1}=J:~\cS\to\psi$
acting in a diagonal way 
on the Fourier coefficients (\ref{4.00}): 
\beq{4.1}
J:~s_{mn}\to 
\wp\!\left[\!\begin{array}{c} m\\n\end{array}\!\right]s_{mn}=\psi_{mn}.
\eq
Here
$$
 \wp\!\left[\!\begin{array}{c} m\\n\end{array}\!\right]=
\wp\left((m+n\tau)\te;\tau\right)\,.
$$

We consider the Euler-Arnold top with the inertia-tensor defined by
$J^{-1}$ (\ref{4.1}).
It has the Hamiltonian
\beq{4.3}
H_\te=\frac{\te^2}{2}\int_{T^2}\cS J(\cS)=
\frac{\te^2}{2}\int_{T^2}\psi J^{-1}(\psi)=
\eq
$$
=\frac{\te^2}{2}\sum_{mn} 
\wp\!\left[\!\begin{array}{c} m\\n\end{array}\!\right]s_{mn}s_{-m,-n}\,,
$$
where the integral is the trace functional (\ref{3.6}).
The equation of motion in the form of the Moyal brackets has the
standard form
\beq{4.4}
\p_t\cS=\te^2\{\cS,J(\cS)\}^*\,,
\eq 
or for the Fourier components 
\beq{4.2}
\p_ts_{mn}=\frac{\te}{\pi}\sum_{j,k}s_{jk}s_{m-j,n-k}\times
\eq
$$
\times\wp\!\left[\!\begin{array}{c} j\\k\end{array}\!\right]
\sin\pi\te(jn-km)\,.
$$

Consider the classical limit $\te\to 0$ of this system
when the Moyal brackets in (\ref{4.4}) pass to the standard Poisson brackets.
In this case we come to the  top on the group SDiff$(T^2)$.
Since
$$
\lim_{\te\to 0}\te^2
\wp\!\left[\begin{array}{c} m\\n\end{array}\right]
=\f1{(m+n\tau)^2}
$$
 the Hamiltonian  (\ref{4.3}) takes the form
\beq{4.3a}
H=\oh\int_{T^2}\psi (\bp)^2\psi\,,
\eq
where $\bp=\f1{\rho}(\tau\p_x-\p_y),~\rho=\tau-\bar\tau$.
The operator $\bp^2$ plays the role of the inertia-tensor.
It replaces the Laplace operator $\Delta\sim\p\bp$ for $\tau=i$
($\p=\f1{\rho}(-\bar{\tau}\p_x+\p_y)$) 
of the standard hydrodynamics.

\section{Integrability of elliptic rotator on ${\cal A}_\te$}
\setcounter{equation}{0}
{\sl 1. The Lax pair}.

We will prove here that the Hamiltonian system of the elliptic rotator
(\ref{4.4}), (\ref{4.2})  have an infinite set of involutive integrals of
motion in addition to the Casimirs (\ref{Ca}). It will follow from the Lax form
\beq{4.5}
\p_tL=[L,M]
\eq
of the equations (\ref{4.4}), (\ref{4.2}).
To define the Lax operator 
we introduce the basic spectral curve
 \beq{bsc}
 E_\tau=\mC/(\mZ+\tau\mZ)\,,
\eq
 with the already defined modular parameter $\tau$.
The Lax operator $L(z)$ is $(1,0)$-form on the spectral parameter 
 $z\in E_\tau$ taking value in the coalgebra $Sin_\te^*$.

Introduce the following functions
\beq{4.6}
\phi(u,z)=\frac{\vth(u+z)\vth'(0)}
{\vth(u)\vth(z)}\,,
\eq
where $\vth(u)$ is the odd theta-function on $E_\tau$,
\beq{A.1a}
\vth(z|\tau)=\bfe\left(\frac
{\tau}{8}
\right)
\sum_{n\in {\bf Z}}(-1)^ne^{\pi i(n(n+1)\tau+2nz)}\,,
\eq
and
\beq{4.7}
\vf\!\left[\!\begin{array}{c} m\\n\end{array}\!\right]\!\!(z)=
\bfe(-n\te z)\phi(-(m+n\tau)\te,z)\,,
\eq
\beq{4.8}
f\!\left[\!\begin{array}{c} m\\n\end{array}\!\right]\!\!(z)=
\eq
$$
=\bfe(-n\te z)\p_u\phi(u,z)|_{u=-(m+n\tau)\te\,}\,.
$$
Then the Lax operator takes the form
\beq{4.9}
L=\sum_{mn}
s_{mn}\vf\!\left[\!\begin{array}{c} m\\n\end{array}\!\right]\!\!(z)T_{mn}\,,
\eq
and
\beq{4.10}
M=\sum_{mn}
s_{mn}f\!\left[\!\begin{array}{c} m\\n\end{array}\!\right]\!\!(z)T_{mn}\,.
\eq
The equivalence of (\ref{4.5}) and (\ref{4.2}) follows from the Calogero
functional equation
$$
\vf\!\left[\!\begin{array}{c} m\\j\end{array}\!\right]
f\!\left[\!\begin{array}{c} j\\n\end{array}\!\right]-
\vf\!\left[\!\begin{array}{c} j\\n\end{array}\!\right]
f\!\left[\!\begin{array}{c} m\\j\end{array}\!\right]=
$$
$$
=\left(
\wp\!\left[\!\begin{array}{c} m\\j\end{array}\!\right]-
\wp\!\left[\!\begin{array}{c} j\\n\end{array}\!\right]
\right)
\vf\!\left[\!\begin{array}{c} m\\n\end{array}\!\right]\,,
$$
and the identity
$$
\vf\!\left[\!\begin{array}{c} m\\n\end{array}\!\right]\!\!(z)
\vf\!\left[\!\begin{array}{c} -m\\-n\end{array}\!\right]\!\!(z)=
\wp(z)-\wp\!\left[\!\begin{array}{c} m\\n\end{array}\!\right]\,.
$$

{\sl 2. Hitchin description.}

It turns out that the elliptic rotator is the Hitchin system \cite{Hi,Ne}
 on $E_\tau$.
It was proved in \cite{LOZ} for the group $\GLN$. Here we replace  $\GLN$
by the infinite-dimensional group $G_\te$. It means that we consider the
infinite rank vector bundle $P$ over $E_\tau$ with the structure group $G_\te$.
The Hitchin systems of infinite rank were introduced in \cite{LOZ,Kr}. 
Here we reproduce the properties of the concrete Lax equation
 as the Hitchin system.
 The Lax operator plays the role of the Higgs
field. It satisfies the Hitchin equation
$$
\bp L=0,~~ Res\,L|_{z=0}=\cS,
$$
with the quasi-periodicity conditions
$$
L(z+1)=QL(z)Q^{-1},
$$
$$
L(z+\tau)=\La L(z)\La^{-1}\,.
$$
It is easy to see that (\ref{4.9}) does satisfy these conditions.

{\sl 3. Integrals of motion.}

 In the Hitchin construction the integrals are defined by means of the
 $(-j,1)$-differentials $\mu_j\in\Om^{(-j,1)}(E_\tau)$.
Let us choose the representatives from $\Om^{(-j,1)}(E_\tau)$  that form 
a basis in the cohomology space
$H^1(E_\tau,\G^j)$, $(\dim H^1=j)$
$$
\mu_j=(\mu_{0,j}\p_z^{j-1}\otimes d\bz,\mu_{2,j}\p_z^{j-1}\otimes d\bz
\ldots
\mu_{j,j}\p_z^j\otimes d\bz)\,.
$$
Let $\chi(z,\bz)$ be a characteristic function of a small neighborhood of
$z=0$. We can choose $\mu_{s,j}$ in the form
$$
\mu_{0,j}\sim \te^j\bp(\bz-z)(1-\chi(z,\bz))\,,
\footnote{$\mu_{0,2}$  is the Beltrami differential}
$$
\beq{4.11a}
\mu_{s,j}=c_{s,j}z^{s-1}\bp\chi(z,\bz)$ for $j>1,\,j\geq s>1,
\eq
$$
c_{s,j}\sim\te^{j-s}\,.
$$
The integrals 
\beq{4.12}
I_{s,j}=\frac{1}{j}\int_{E_\tau}\int_{T^2}(L^j)\mu_{s,j}
\eq
are well defined and generate the infinite set of conservation laws.
 Here we integrate over the
basic spectral curve $E_\tau$ (\ref{bsc}) and NCT ${\cal A}_\te$.
The conservation laws can be extracted from the expansion of the
elliptic function
$$
\frac{1}{j}\int_{T^2} L^j(z)=I_{0,j}+\sum_{r=2}^{j}I_{r,j}\wp^{(r-2)}(z),~
(j=2,\ldots)\,.
$$
In particular,
$$
\int_{T^2} L^2(z)=I_{0,2}+\wp(z)\int_{T^2}\cS^2,~H=I_{0,2}\,.
$$
Note that 
$$
I_{j,j}\sim C_j=\int_{T^2}\cS^j
$$
 are the Casimirs (\ref{Ca}).

Consider, for example, the integrals, that
 have the third order in the field $\cS$ .
Let 
$$
E_1(u)=\p_u\log\vth(u)\,,
$$
be the first Eisenstein function.\footnote{It is related 
to the Weierstrass zeta-function
as $E_1(u)=\ze(u)-2\ze(1/2)u$.}
The second  Eisenstein function is
 $$
E_2(u)=\wp(u)+2\ze(1/2)\,.
$$
For the functions $\phi(u,z)$ (\ref{4.6}) we have the following relation.
If $u_1+u_2+u_3=0$ then
\beq{4.15}
\phi(u_1,z)\phi(u_2,z)\phi(u_3,z)=-\oh E'_2(z)+
\eq
$$
E_2(z)\left(E_1(u_1)+E_1(u_2)+E_1(u_3)\right)+
$$
$$
E_2(u_3)\left(E_1(u_1)+E_1(u_2)+E_1(u_3)\right)
-\oh E'_2(u_3)\,.
$$
This relation allows to calculate $\int_{T^2} L^3(z)$.
Let
$$
E_2\ar{m}{n}=E_2((m+n\tau)\te),
$$
$$
E_1\ar{m}{n}=E_1((m+n\tau)\te)\,.
$$
Then in terms of the Fourier modes $\cS=\{s_{mn}\}$ the integrals take the form
\beq{4.16}
I_{1,3}=\frac{\te^3}{3}\sum_{\sum m_j=\sum n_j=0}
\prod_{j=1}^3s_{m_jn_j}\times
\eq
$$
\left(
E_2\ar{m_3}{n_3}\sum_jE_1\ar{m_j}{n_j}-\oh E'_2\ar{m_3}{n_3}
\right)\,,
$$
\beq{4.17}
I_{2,3}=-\frac{\te}{6}\sum_{\sum m_j=\sum n_j=0}
\prod_{j=1}^3s_{m_jn_j}\times
\eq
$$
\times\sum_jE_1\ar{m_j}{n_j}\,.
$$

The integrals (\ref{4.12}) allow to write the hierarchy of the commuting
flows
\beq{4.13}
\p_{s,j}\cS=\{\nabla I_{s,j},\cS\}^*~~~(\p_{s,j}=\p_{t_{s,j}})\,.
\eq
They have the Lax representation with $L$ (\ref{4.9})
\beq{4.14}
\p_{s,jL}=[L,M_{s,j}]:=\{L,M_{s,j}\}^*\,,
\eq
 and 
$M_{s,j}$ is partly fixed by the equation
$$
\bp M_{s,j}=-L(z)^{j-1}\mu_{s,j}\,.
$$

{\sl 4. Classical limit}.

In the limit $\te\to 0$ the Lax hierarchy (\ref{4.14}) is replaced by the
 classical equations
$$
\p_{s,j}L^{(-1)}=\{L^{(-1)},M^{(0)}_{s,j}\}\,,
$$
where
$$
L^{(-1)}=\frac{s_{mn}}{(m+n\tau)}
$$
 and $M_{s,j}=M^{(0)}_{s,j}+O(\te)$.
 The integrals of motion in this limit survive.
We already pointed the form of the Hamiltonian $H=I_{0,2}$ (\ref{4.3a}).
The third order integrals take the forms
$$
I_{1,3}=\frac{1}{3}\sum_{\sum m_j=\sum n_j=0}
\prod_{j=1}^3s_{m_jn_j}\times
$$
$$
\left(
\f1{(m_3+n_3\tau)^2}\sum_j\f1{m_j+n_j\tau}- \f1{2(m_3+n_3\tau)^3}
\right)\,,
$$
$$
I_{2,3}=-\f1{6}\sum_{\sum m_j=\sum n_j=0}
\prod_{j=1}^3s_{m_jn_j}\times
$$
$$
\times\sum_j\f1{m_j+n_j\tau}\,.
$$

It turns out that  the classical limit in some sense is the same as
the rational limit of the basic spectral curve $E_\tau$.
Let $\om_1,\om_2$ be two half-periods of $E_\tau,~\tau=\om_2/\om_1$.
The rational limit means that $\om_1,~\om_2\to\infty$.
In this case the Weierstrass function is degenerate as
$$
\wp(u)\to \f1{u^2}\,.
$$
Consider the double limit $\te\to 0$, $\om_1,~\om_2\to\infty$
 such that
$$
\lim\om_1\te=1,~~\lim\om_2\te=\tau
$$
It follows from the definition of the
 Weierstrass function $\wp(u;\om_1,\om_2)$ that in this limit
$$
\wp\!\left[\!\begin{array}{c} m\\n\end{array}\!\right]\to
\f1{(m+n\tau)^2}\,.
$$
Rescale now the Hamiltonian  (\ref{4.3}) by the dropping the $\te^2$
multiplier. The new Hamiltonian in the double limit takes the form
(\ref{4.3a}).

\section{Conclusion}
There are three related subjects that are not covered here. The first
will be elucidated in separated publications.

$\bullet$
 The elliptic rotator on $\GLN$ can be transformed to the
elliptic Calogero-Moser system (ECM) by the symplectic Hecke correspondence \cite{LOZ}.
For general orbits we obtain ECM system with spin. The spin degrees
of freedom disappear for the most degenerated orbits. The similar approach
can be developed for the elliptic rotator on ${\cal A}_\te$. In other words,
there should exists a symplectic Hecke correspondence between
 a special thermodynamical limit of ECM and the elliptic rotator on ${\cal A}_\te$.
 The general  
thermodynamical limit of the ECM system
leads to ECM system related  to GL$(\infty)$,
 but here we consider the NCT subgroup and it means that the limit is special.
For $\te=p/N$  after the factorization over the ideal
(\ref{I1}) we come back to the ECM system for $\GLN$. 
I plan to describe  
ECM system for ${\cal A_\te}$  and its symplectic (Hecke) transformation 
to the elliptic EAT for ${\cal A_\te}$.

$\bullet$
The EAT can be considered on the central extended algebra
$\widehat{Sin}_\te$ (\ref{3.13}). For the hydrodynamics $(\gJ=\D)$ this case
was investigated in Ref.~\cite{Ze}. Incorporating of the central
charge  drastically changes the dynamics of EAT. In particular,
the integrals (\ref{4.12}) are no longer the conserved quantities, as well as
 the Hamiltonian (\ref{4.3}).
 For ECM system related to $\GLN$ this construction
leads to the two-dimensional ECM integrable 
field theory \cite{LOZ,Kr}. 
In the group theoretical terms it means that we pass
from $\GLN$ to the central extended
loop group $\hat{L}(\GLN)$.

The same symplectic transformation that maps ECM system 
to the elliptic EAT is working in the
two-dimensional case. In particular, the two-dimensional ECM system for $N=2$ 
is mapped to the Landau-Lifshitz equation \cite{LOZ}.
In the limit $N\to\infty$ we obtain the symplectic map from the
thermodynamical limit of the ECM field theory to the central extended 
elliptic rotator. The both cases can be described as the systems related
to $\widehat{Sin}_\te$. When $\te=p/N$ after the factorizing over the ideal 
(\ref{I2}) we come back to the systems considered in \cite{LOZ,Kr}.
  
$\bullet$
Two different tori are incorporated in our construction -
NCT ${\cal A}_\te$ and the basic spectral curve $E_\tau$.
In the classical limit they become dual. It seems natural to replace
 $E_\tau$ on another NCT  ${\cal A}_{\te'}$. In a general setting it means
a generalization of  the Hitchin systems to the non-commutative case.
 One attempt in this direction was done in Ref.\cite{T2}.
As we have seen, the theta-functions with characteristics are 
the building blocks
for the Lax operators. 
 The analogs of theta-functions on NCT are
known  \cite{We,Ma,Sch}. We can hope that they will play the essential
role in the construction of the Hitchin type systems 
in the non-commutative situation.

\section*{Acknowledgments}

The hospitality of the Max Planck Institute of Mathematics, Bonn,
during the preparation of this article is gratefully acknowledged.
I am grateful to D.Lebedev and A.Vershik for fruitful discussions and to 
A.Tsygvintsev, who pointed me the paper \cite{Z}.
The work is supported in part  by grants
RFBR-00-02-16530, 00-15-96557 for support of scientific schools,
and INTAS-99-01782.

\end{document}